\documentclass[a4paper,12pt]{article}
\usepackage{epsfig}
\usepackage{amsfonts,amsmath,amssymb}
\usepackage{mathtools}
\usepackage{a4wide}
\usepackage{bm}
\usepackage{braket}
\usepackage{slashed}
\usepackage{subfig}
\usepackage{booktabs}
\usepackage{multirow}
\usepackage{color}
\newcommand{\mysmall}[1]{\rm \scriptscriptstyle #1} 
\usepackage{cite}

\newcommand{\pks}{\bm p_{K^*}}
\newcommand{\mks}{m_{K^*}}

\begin{document}
\unitlength1cm
\begin{titlepage}
\vspace*{-1cm}
\begin{flushright}
SI-HEP-2018-12\\
QFET-2018-06
\end{flushright}
\vskip 3.5cm

\begin{center}
{\Large\bf \boldmath On the decays $B \to K^{(*)} + $ leptonium}
\vskip 1.cm
{\large  Matteo~Fael} and {\large Thomas~Mannel}
\vskip .7cm
{\it
Theoretische Physik I, Universit\"at Siegen, 57068 Siegen, Germany}
\\[2mm]

\end{center}
\vskip 2cm

\begin{abstract}
\noindent We determine the rates of the $B$ meson decays into a $K^{(*)}$ and an $\ell^+ \ell^-$ bound state, the leptonium, where $\ell = e,\mu,\tau$.
  The two spin states of the leptonium, the spin singlet and the spin triplet, couple to the axial current and to the vector current,  respectively, thus probing different helicity structures of the underlying $b\to s \ell^+\ell^-$ effective Hamiltonian. 
  Since ortho- and para-leptonia have different decay modes, a distinction between the two is relatively easy and these decays may become a cross check for the results of lepton-flavour-violation searches obtained with free leptons. 
  We find that some of the decays involving muon and tau have a branching ratio of the order of $10^{-13}$ and they may become accessible at the LHCb with 50 fb$^{-1}$  of integrated luminosity.
  In addition, since the tau-pair threshold lies right between the $J/\psi$ and the $\psi(2S)$ resonances, we estimate the charm-loop contribution to the decays  $B \to K^{(*)} + $tauonium.
\end{abstract}
\end{titlepage}
\newpage

\section{Introduction}
Currently the rare decays mediated by the transition $b \to s \ell^+ \ell^-$ exhibit small tensions 
when comparing the data with the predictions of the Standard Model (SM)~\cite{Aaij:2014ora,Aaij:2017vbb}. In particular, there is 
a (still not significant) anomaly indicating lepton universality violation (LUV). 
Global fits to the available $b\to s \ell^+\ell^-$ data  for the Wilson coefficients $C_{7\gamma}^{(')},C_{9}^{(')}$ and $C_{10}^{(')}$ indicate a large deviation in $C_9$, $\Delta C_9 \simeq -1$, with the SM hypothesis disfavored at the level of 4-5$\sigma$~\cite{Capdevila:2017bsm,Altmannshofer:2017yso,Geng:2017svp,Ciuchini:2017mik}.

The measurements of $B\to K^{(*)} \ell^+ \ell^-$ have reached a sufficient precision to study the rates in bins of the leptonic invariant mass, in particular also close to the threshold for $\mu^+ \mu^-$. 
In this region  threshold effects from QED may play a role~\cite{Bordone:2016gaq}. One of these effects is the formation of leptonia, such as positronium, dimuonium or even tauonium, which has motivated us to study the decays $B \to K^{(*)} + $ leptonium. 

Already a naive power counting of the electromagnetic coupling $\alpha$ shows that these effects --- for instance dimuonium decaying into $e^+e^-$ --- cannot explain the observed hints of LUV in the lowest bin of the leptonic invariant mass.
Nevertheless, these decays may become an interesting cross check for the results obtained by using states with ``open'' leptons.

In the following sections we compute the rates for the decays $B \to K^{(*)} + $ leptonium in terms of the wave functions at the origin of the leptonia. 
Since to leading order in $\alpha$ only the $S$ wave leptonia matter, we only have to consider the case of ortho- and para-leptonia in $S$ wave states. Since ortho- and para-leptonia have different decay modes (e.g. a two-photon decay versus a 
three-photon decay) a distinction between the two is relatively easy. On the other hand, 
ortho- and para-leptonia probe different helicity structures of the underlying interaction, allowing us 
an independent test. 

In addition, the threshold for the tauonuim is much higher than the one for the light leptonia and lies 
close to some of the charm resonances. Thus the prediction for the tauonium case is more difficult, 
but the contribution of the  charm-loops appears to be not too big. 

Some of the branching ratios of $B \to K^{(*)} + $leptonium involving muons and taus turn out to be of the $O(10^{-13})$.
In the $pp$ collisions at the LHC the $\bar{b}b$ cross section at $\sqrt{s}=14$~TeV is about $500\mu$b, which corresponds to the production of $10^{12} \, \bar{b}b$ pair in a standard year  of running ($10^7$ s) at the LHCb operating luminosity of $2 \times 10^{32} \, \text{cm}^{-2} \, \text{sec}^{-1}$~\cite{Alves:2008zz}.
Therefore, a subset of the $B \to K^{(*)} + $leptonium decays may become accessible at the LHCb; others however are too small even for the high-luminosity run.

In the next section we discuss the properties of the leptonia needed for our purpose; in section~\ref{sec:oniumdecayconstants} we give the necessary matrix elements of the leptonia states.
Our results for branching ratios stemming from $O_{7\gamma}, O_9$ and $O_{10}$ are presented in section~\ref{sec:branchingratios}, while in section~\ref{sec:charmloop} we will discuss the charm-loop contribution coming from the four-quark operators $O_1$ and $O_2$. Conclusions are drawn in section~\ref{sec:conclusions}.

\section{\boldmath The $\ell^+\ell^-$ Bound States}
\label{sec:onium}
In this section we recall some basic properties of the bound states of two leptons. 
We will refer to generic bound state of $\ell^+\ell^-$ as \emph{leptonium}. 
Positronium  and tauonium  are the bound states of $e^+e^-$ and $\tau^+ \tau^-$, respectively, while the bound 
state of a muon and an anti-muon is called dimuonium or true muonium.\footnote{The term ``muonium'' is often also used for a state 
composed of a muon and an electron. Since we do not consider such a state in this paper, there cannot be any confusion.}
Positronium was discovered already in 1951~\cite{Deutsch:1951zza}, however dimuonium  and tauonium have not been observed yet since their very narrow widths cannot be resolved by $e^+e^-$ colliders because of beam energy spread. 
A low-energy electron-positron collider to search and study dimuonium  is currently under consideration at the Budker Institute for Nuclear Physics~\cite{Bogomyagkov:2017uul}.

For the leptonia we make use of the spectroscopic notation known from atomic physics 
\begin{equation}
  n^{2S+1} L_J,
\end{equation}
where $n$ is the principal quantum number, $S$ is the total spin quantum number --- it can be 0 or 1 --- $L$ is the orbital angular momentum quantum number with $L=S,P,D,\dots$ for $L=0,1,2,\dots$, and $J$ is the total angular momentum quantum number.
Using only the coulombic interaction between the two leptons, 
the leptonium mass $M_n$ differs from the sum of two lepton masses by the small binding energy $\mathcal{E}_n$:
\begin{equation}
  M_n=2m-\mathcal{E}_n, \mbox{ with } \mathcal{E}_n = \frac{m\alpha^2}{4n^2},
\end{equation}
where $m$ denotes the mass of the lepton $\ell$.
In the rest frame of the leptonium the two leptons are non-relativistic since their three-momenta are of the order $\alpha m \ll m$.

To leading order in $\alpha$ only $S$ wave leptonia (i.e.\ states with zero angular momentum) can contribute, 
since the lepton currents appearing in the effective Hamiltonian for the $b \to s \ell^+ \ell^-$ interaction has only local 
leptonic currents, which means that only the wave function at the origin of the leptonium matters.   
The position-space Schr\"{o}dinger wave function is different from zero at the origin only for an $L=0$ state and is given by
\begin{equation}
  |\phi_n(0)|^2 = \frac{(m\alpha)^3}{8\pi n^3}.
\end{equation}
There are two types of bound states depending on the $\ell^+\ell^-$ spin sum:  the spin-0 singlet --- $^1 S_0$ in 
spectroscopic notation --- and the spin-1 triplet --- $^3 S_1$. We borrow the notations from the positronim and call 
the  $n^1 S_0$ bound states para-leptonium and the $n^3 S_1$ states ortho-leptonium. 

The leptonium states are unstable. For the  positronum ground state the only decay mode is by annihilation of the $\ell^+\ell^-$ pair into photons, 
while the muonium and the tauonium can decay also through a pair creation of lighter leptons and by the weak decay of the lepton.   
A spin-0 state can annihilate only into an even number of photons while a spin-1 triplet into an odd number of photons. 
This is a consequence of charge conjugation: the $C$-parity of the bound state is given by $(-1)^{S+L}$ while for a system of $n$ photons it is $(-1)^n$.
This means for example that the ground-state para-positronium decays primarily into two photons while the ortho-positronium into three photons, 
since the decays into one (off-shell) photon is forbidden by energy conservation.
The decay rate of a spin singlet into two photons is (see e.g.~\cite{Itzykson:1980rh,Berestetsky:1982aq})
\begin{equation}
  \Gamma (\text{para-leptonium} \to 2 \gamma) = \frac{m\alpha^5}{2n^3},
\end{equation}
while the decay rate of a spin triplet into three photons is
\begin{equation}
\Gamma (\text{ortho-leptonium} \to 3 \gamma) = \frac{2 (\pi^2-9) \alpha^6 m}{9\pi n^3},
\end{equation}
which is suppressed by an extra power  of $\alpha$ with respect to the spin singlet. This gives to the ground-state ortho-positronium a lifetime longer than the para-positronium.\footnote{Higher order corrections to the positronium lifetime were calculated in~\cite{Kniehl:2009pg,Adkins:2015wya,Adkins:2003eh,Adkins:2001zz,Adkins:2000fg,Penin:2003jz,Kniehl:2000dh,Karshenboim:2005iy,Hill:2000qi,Melnikov:2000fi,Czarnecki:1999gv,Czarnecki:1999ci}.}

Turning to the muonium and tauonuim case, we note that the ortho-muonium ground state dominantly decays into an $e^+ e^-$ pair via virtual single-photon annihilation; likewise the ortho-tauonium decays dominantly into $e^+e^-$ or $\mu^+\mu^-$, approximatively with the same rate, or into hadrons 
via $q$ $\bar{q}$ channels.  The rate of this decay mode is of the same order in $\alpha$  as the one of the of the 
spin singlet radiative decay and given by\footnote{Higher order corrections to the dimuonium lifetime were calculated in~\cite{Malenfant:1987tm,Jentschura:1997tv,Karshenboim:1998we}}
\begin{align}
  \Gamma (\text{ortho-muonium} \to e^+ e^-) &= \frac{\alpha^5 m_\mu}{6n^3},  \\
  \Gamma (\text{ortho-tauonium} \to e^+ e^-) &= \Gamma (\text{ortho-tauonium} \to \mu^+ \mu^-) =
  \frac{\alpha^5 m_\tau}{6n^3} \\
  \Gamma (\text{ortho-tauonium} \to \text{hadrons}) &= 
  \frac{\alpha^5 m_\tau}{6n^3} R^\text{had}(4m_\tau^2) 
  \simeq \frac{\alpha^5 m_\tau}{6n^3} \sum_{q=u,d,s}  N_c \, Q_q^2 \, , 
  \label{eqn:ontoellell}
\end{align}
where $R^\text{had}(s) \equiv \sigma(e^+e^- \to \gamma^* \to \text{hadrons})/\sigma(e^+e^- \to \gamma^* \to \mu^+\mu^-)$, $N_c=3$ and $Q_q$ is the charge of the quark $q$ in units of the electron charge.

In addition for $\ell=\mu, \tau$ the $\ell^+\ell^-$ atom can break  up because of the weak decay of one of the constituent leptons. 
Since the decay of both lepton disintegrates the leptonium, the decay width is  approximately given by~\cite{Perl:1993sk}:
\begin{equation}
  \Gamma(\mbox{leptonium, } \ell\,  \text{decay}) = {2 \over \tau_\ell},
\end{equation}
with $\tau_\ell$ the lifetime of the muon or the tau. However, the rate for the weak decays of the leptons scales with the fifth power of the mass, while the rates from the annihilation processes scale as  $m$. 
Therefore,  for some value of the mass $m$, the lifetime of the bound state becomes longer than the lifetime of the 
lepton, so that the leptonium decays before the $\ell^+ \ell^-$ pair can annihilate.
For the muonium we have 
\begin{align}
  \Gamma(\mbox{muonium}, \,\mu \, \text{decay}) &= 
  6.0 \times 10^{-10} \, \text{eV},  \\
  \Gamma (\mbox{para-muonium}  \to 2 \gamma) &=
  {1.10 \times 10^{-3} \, \text{eV} \over n^3},  \\
  \Gamma (\mbox{otho-muonium}  \to e^+e^-) &=
  { 3.64 \times 10^{-4} \, \text{eV} \over n^3} \, , 
\end{align}
thus the comparison of the three partial widths tells us that the $\mu^+\mu^-$ annihilation is the dominant mode and the muon lifetime is much longer than the muonium one.

For the tau lepton this is different: 
\begin{align}
  \Gamma(\mbox{tauonium}, \,\tau \, \text{decay}) &= 
  4.53 \times 10^{-3} \, \text{eV},  \\
  \Gamma (\text{para-tauonium} \to 2 \gamma) &=
  {1.84 \times 10^{-2} \, \text{eV} \over n^3},  \\
  \Gamma (\text{ortho-tauonium} \to e^+e^-, \, \mu^+\mu^-) &=
  { 6.13 \times 10^{-3} \, \text{eV} \over n^3}.
\end{align}
In this case only the ground state $n=1$ decays before the decay of the tau. 

In addition, there are of course also electromagnetic transitions within the leptonium system from an upper energy level to a lower one. An excited $^{2S+1}S_1$ leptonium state can decay via electric dipole transition, which conserves the spin quantum number,  to a $^{2S+1}P_J$ state and a soft photon with energy of the order of $\mathcal{E}_n \sim m \alpha^2$. 
Magnetic transitions with $\Delta S \neq 0$ are much more suppressed.
Since the $C$-parity of the $\ell^+\ell^-$ system is given by $(-1)^{L+S}$, the $P$ state has now opposite $C$-parity with respect to the initial $S$ state.
This can potentially spoil the possibility to distinguish between para- and ortho-leptonium.
However the annihilation probability of a $P$ state is proportional to the derivative of the wave function at the origin which leads to an extra $\alpha$ suppression in the decay rate compared to the $S$ wave case.
Therefore such $P$ state decays primarily, after one or more transitions, again into a state in $S$ wave with lower $n$ but conserving the original spin quantum number.

\section{The Leptonium Decay Constants}
\label{sec:oniumdecayconstants}
In this section we calculate the matrix element 
\begin{equation}
\bra{0} j_{V,A}^\mu (0) \ket{n, q, \sigma}, 
\label{eqn:decayconstantdefinitions}
\end{equation}
that will enter in the decay amplitude of $B\to K^{(*)} +$ leptonium (see e.g. the review~\cite{Novikov:1977dq}). 
Let us denote with $\ket{n, q, \sigma}$ the leptonium one-particle state with principal quantum number $n$ and four momentum $q = (E_{n,\bm q}, \bm q)$, with $E_{n,\bm q} = \sqrt{\bm q^2+M_n^2}$. The index $\sigma$ indicates one of the four spin states: three spin-1 states (triplet) and one spin-0 state (singlet).
The leptonium one-particle state is normalized in a Lorenz invariant way: 
$\langle n,q,\sigma | n',q',\sigma'\rangle = 
(2\pi)^3 \, 2 E_{n,\bm{q}} \, \delta^{(3)}(\bm q - \bm q') \, \delta_{nn'} \delta_{\sigma \sigma'}$.
The vector and axial currents are $j^\mu_V(x) = \overline{\ell}(x) \gamma^\mu \ell(x)$ and $j_A^\mu (x) =\overline{\ell}(x) \gamma^\mu \gamma_5 \ell(x)$, respectively.

In a non-relativistic picture --- which is well justified for the leptonia since $| \bm k| \sim \mathcal{O}(\alpha m) \ll m$ --- the one-particle state can be expressed as linear superposition of free $\ell^+$ and $\ell^-$ states with three-momenta $\bm k_\pm$, respectively, and energies $E_\pm=\sqrt{\bm k_\pm^2+m^2}$:
\begin{equation}
\ket{n,q,\sigma} = 
\int \frac{d^3 k}{(2\pi)^3}
\sqrt{\frac{2E_{n,\bm q}}{2 E_+ 2E_-}}
\tilde{\phi}_{n,\bm q}(\bm k) 
\ket{k_+,k_-,\sigma},
\label{eqn:oniumtofree}
\end{equation}
with $\bm q = \bm k_+ + \bm k_-$ and $\bm k = (\bm k_+-\bm k_-)/2$. 
This superposition is weighted by the momentum-space Coulomb wave function $\tilde{\phi}_{n,\bm q}(\bm k)$, which gives the amplitude for finding a particular value of $\bm k$ for a leptonium state $n$ with total momentum $\bm q$; the wave function $\tilde{\phi}_{n,\bm q}(\bm k)$ fulfils the normalization condition $\int \frac{d^3 k}{(2\pi)^3} |\tilde{\phi}(k)|^2 = 1$.
To evaluate the matrix element~\eqref{eqn:decayconstantdefinitions} it is sufficient to consider only the state with $\bm q = \bm 0$, $\bra{0} j^\mu_{V,A}(0) \ket{n,\bm 0,\sigma}$, since all other states with $\bm q \neq 0 $ can be obtained via a boost:
\begin{equation}
\bra{0} j^\mu_{V,A}(0) \ket{n,\bm q,\sigma} = 
\bra{0} \hat U^\dagger(\bm q) j^\mu_{V,A}(0) \hat U(\bm q) \ket{n,\bm 0,\sigma} =
\Lambda^\mu \,_\alpha(\bm q) \bra{0} j^\alpha_{V,A}(0) \ket{n,\bm 0,\sigma},
\label{eqn:boost}
\end{equation}
where $\hat U(\bm q)$ is the unitary operator that boosts the state $\ket{n,\bm 0,\sigma}$, where the momentum is $q^* = (M_n,\bm 0)$, into the state $\ket{n,\bm q,\sigma}$ and $\Lambda^\mu \,_\alpha(\bm q)$ is the corresponding Lorentz matrix.
Thanks to the expansion~\eqref{eqn:oniumtofree}, the evaluation of the matrix element involving the one-particle state is reduced to the evaluation of 
\begin{equation}
\bra{0} j_{V,A}^\mu(0) \ket{k_+,k_-,\sigma} =
\overline{v}_\sigma (k_+) j_{V,A}^\mu u_\sigma (k_-),
\label{eqn:matelem}
\end{equation}
where $k_\pm$ can be taken in the rest frame of the leptonium and the subscript ``$\sigma$'' refers to the fact that the free $\ell^+$ and $\ell^-$ states must properly combine into the leptonium spin state $\sigma$.
We can now employ the expressions for the spinors in the non-relativistic approximation:
\begin{align}
u_\sigma &= \sqrt{m} 
\begin{pmatrix}
\xi_\sigma \\
\xi_\sigma
\end{pmatrix}, &
v_\sigma &= \sqrt{m} 
\begin{pmatrix}
\eta_\sigma \\
-\eta_\sigma
\end{pmatrix},
\end{align}
where $\xi$ and $\eta$ are two-component spinors normalized to unity. The expression in~\eqref{eqn:matelem} can be written as a trace of $2\times2$ matrix chain, indeed the spinor product $\xi_\sigma \eta_\sigma^\dagger$ can be replaced for a spin-1 state by
\begin{equation}
\xi_\sigma \eta_\sigma^\dagger
\to \frac{\bm \epsilon_\sigma \cdot \bm \sigma}{\sqrt{2}},
\end{equation}
where $\bm \sigma= (\sigma_1,\sigma_2,\sigma_3)$ are the three Pauli matrices and $\bm \epsilon_\sigma$ are unit polarization vectors: for $\bm q$ along the $z$-axis, $\bm \epsilon_\sigma = (0,0,1)$ corresponds to the longitudinal polarization while $\bm \epsilon_\sigma = (1,i,0)$ and $\bm \epsilon_\sigma = (1,-i,0)$ to the transverse polarizations.
In the same way, the spin-0 state is given by the replacement
\begin{equation}
\xi_\sigma \eta_\sigma^\dagger
\to \frac{\bm 1}{\sqrt{2}}.
\label{eqn:repsinglet}
\end{equation}
We can now compute the matrix element containing the vector current:
\begin{equation}
\bra{0} j_{V}^\mu(0) \ket{k_+,k_-,\sigma} =
m \eta^\dagger_\sigma (\bar \sigma^\mu-\sigma^\mu)\xi_\sigma = 
\frac{m}{\sqrt{2}}\epsilon_\nu(q^*)
\mathrm{Tr} \left(\sigma^\nu \bar \sigma^\mu- \bar \sigma^\nu \sigma^\mu\right) =
-\frac{4 m}{\sqrt{2}} \epsilon_\sigma^\mu(q^*),
\label{eqn:matV}
\end{equation}
where we introduced the four-vectors $\epsilon^\mu_\sigma (q^*) = (0,\bm \epsilon_\sigma)$, $\sigma^\mu = ( 1, \bm \sigma)$ and $\bar \sigma^\mu = ( 1, -\bm \sigma)$.
For the axial current we obtain
\begin{equation}
\bra{0} j_{A}^\mu(0) \ket{k_+,k_-,\sigma} =
-m \, \eta^\dagger_\sigma (\bar \sigma^\mu+\sigma^\mu)\xi_\sigma = 
-\frac{m}{\sqrt{2}}
\mathrm{Tr} \left( \bar \sigma^\mu+\sigma^\mu\right) =
-\frac{2}{\sqrt{2}} q^{*\mu}.
\label{eqn:matA}
\end{equation}
In the last expression the trace vanishes if $\mu=1,2,3$ while for $\mu=0$ it is equal to four, i.e.\ the trace must be proportional to the leptonium momentum in its rest frame $q^*=(M_n,\bm 0)$, with $M_n \sim 2 m$.  
The vector current $j_V^\mu$ couples only to the spin-1 states, in fact for a singlet the replacement in eq.~\eqref{eqn:repsinglet} yields a vanishing trace. In the same way, one verifies that the axial current couples only to the spin-0 state and not to the triplet.  
  
The expressions~\eqref{eqn:matV} and \eqref{eqn:matA} can be inserted in eq.~\eqref{eqn:oniumtofree}, and the integral can be done by using the non-relativistic approximation for the energies in the leptonium rest frame, $E_\pm \sim M_n/2 \sim m$ and $E_{n,\bm q} \sim M_n$. After performing the boost in eq.~\eqref{eqn:boost}, we obtain the following expressions of the matrix elements:
\begin{align}
  \bra{0} j_V^\mu (0) \ket{n,q,\sigma} &=
  \begin{dcases}
    -2 \sqrt{M_n} \, \phi_{n,\bm 0} (0) \, \epsilon^\mu_\sigma(q) &
    \mbox{for spin-1 state}, \\
    0 & 
    \mbox{for spin-0 state}, 
  \end{dcases} 
  \label{eqn:matelemvector}
  \\[5pt]
  \bra{0} j_A^\mu (0) \ket{n,q,\sigma} &= 
  \begin{dcases}
    0 &
    \mbox{for spin-1 state}, \\
    -\frac{2 }{\sqrt{M_n}} \, \phi_{n,\bm 0} (0) \, q^\mu & 
    \mbox{for spin-0 state}, 
  \end{dcases}
  \label{eqn:matelemaxial}
\end{align}
where $\phi_{n,\bm 0}(0)$ is the position-space wave function at the origin in the rest frame of the leptonium.
The rate $\Gamma(\text{ortho-leptonium} \to e^+ e^-)$ in~\eqref{eqn:ontoellell} can be calculated from the amplitude in eq.~\eqref{eqn:matelemvector}.
The matrix elements can be cast also in terms of decay constants, as for pseudoscalar and vector mesons~\cite{Pivovarov:1991vb}:
\begin{align}
  \bra{0} j_V^\mu(0) \ket{\text{spin-1}} &= M_n f_V \, \epsilon_\mu(q), &
  \bra{0} j_A^\mu(0) \ket{\text{spin-0}} &=  f_A \, q^\mu, 
\end{align}
with
\begin{equation}
f_V = f_A = - \frac{2}{\sqrt{M_n}} \phi_{n,\bm 0}(0). 
\end{equation}

\section{Branching Ratios}
\label{sec:branchingratios}
With the input of the leptonium-decay constants we can now evaluate the branching ratio $B \to K^{(*)} +$ leptonium. 
The decay amplitude of $B \to K^{(*)} + $ leptonium can be written as:
\begin{align}
&  \mathcal{A}(B\to K^{(*)}+ \, \text{leptonuim}) = 
  -\bra{K^{(*)}(p) \, \text{leptonium} (q) }
  H_{\mysmall eff}
  \ket{B(p+q)}  \\  \notag
  & \qquad =\frac{G_F}{\sqrt{2}} \frac{\alpha}{\pi} V_{tb} V_{ts}^*
  \Big[ 
    \bra{n,q,\sigma} \bar \ell \gamma_\mu \ell \ket{0}
    \mathcal{H}_V^\mu(p,q)
    +
    \bra{n,q,\sigma} \bar \ell \gamma_\mu \gamma_5 \ell \ket{0}
    \mathcal{H}_A^\mu(p,q)
  \Big],
\end{align}
where $H_{\mysmall eff}$ is the effective Hamiltonian for $b\to s \ell^+\ell^-$ transitions while $\mathcal{H}_V^\mu(p,q)$ and $\mathcal{H}_A^\mu(p,q)$ are the parts of the hadronic matrix element which are 
contracted with the leptonic vector current and axial current, respectively; their explicit expressions can be obtained from the $B\to K^{(*)} \ell^+ \ell^-$ amplitudes reported in the appendix. 

We have calculated the two leptonic matrix elements in the last section, where we have seen that the 
vector current couples only to the ortho-leptonium, while the axial current couples exclusively to the para-leptonium. 
In this sense, the ortho- and the para-leptonium have to be considered as two different particles with different decay modes, 
and this allows us to probe different aspects of the underlying interaction by looking at the two kinds of leptonia, which 
are easily distinguished by their different decay modes. To this end, we 
will distinguish two decay rates:
\begin{equation}
  \Gamma(B\to K^{(*)} + \text{para-leptonium}) \quad \mbox{ and } \quad
  \Gamma(B\to K^{(*)} + \text{ortho-leptonium}).
\end{equation}

While the decay modes are quite different,  the masses of the leptonia differ only by a few keV (for $\ell=\mu,\tau$), which cannot be resolved by
experiments such as LHCb.  Thus we identify all masses to be $M_n = 2 m$ and compute the sum over all 
$S$ wave leptonia; since in this approximation the  principal quantum number $n$ appears only in the squared wave function at the origin,
\begin{equation}
  |\phi_{n,\bm 0}(0) |^2 = \frac{(m\alpha)^3}{8\pi n^3} \, ,
\end{equation}
we pick up a factor  
$$
\sum_{n=1}^\infty \frac{1}{n^3}= \zeta(3) \, . 
$$
However this summation is performed only for the light leptonia; for the tauonium only the $n = 1 $ state can be considered, 
all higher tauonia will decay through the weak decay of the $\tau$ before they actually form. 
 
For the decay $B \to K +$ ortho-leptonium we obtain the rate
\begin{multline}
  \Gamma(B \to K + \text{ortho-leptonium}) = \\
  \frac{G_F^2 |\bm p_K|^3 m^2}{64 \pi^4}
  \alpha^5 \zeta(3) |V_{tb} V_{ts}^*|^2 
   \left\vert C_9 f^+(4m^2) 
   + C_{7\gamma}^{\mysmall eff} \frac{2(m_b+m_s)}{m_K+m_B}
   f^T(4m^2)\right\vert^2 \,,
  \label{eqn:rateBtoKspin1}
\end{multline}
where we have summed over the three spin states of the ortho-leptonium. Note that for the ortho-tauonium the factor 
$\zeta(3)$ has to be replaced by $1$. 

For $B \to K +$ para-leptonium we get 
\begin{multline}
  \Gamma(B \to K + \text{para-leptonium}) = 
  \frac{G_F^2 m_B^2 |\bm p_K|m^2}{16 \pi^4}
  \alpha^5 \zeta(3) |V_{tb} V_{ts}^*|^2 
  \\ 
  \times \Bigg\{\frac{|\mathcal{L}_1|^2}{16}
   \left( 1-\frac{m_K^2+4m^2}{m_B^2} \right)^2 
 +4|\mathcal{L}_2|^2 \frac{m^4}{m_B^4} 
   +\mathrm{Re}\left( \mathcal{L}_1 \mathcal{L}_2^* \right)
   \frac{m^2}{m_B^2}\left( 1-\frac{m_K^2+4m^2}{m_B^2} \right)
 \Bigg\} \, ,
  \label{eqn:rateBtoKspin0}
\end{multline}
where $\mathcal{L}_i = \mathcal{L}_i(q^2 = 4m^2)$, with
\begin{align}
  \mathcal{L}_1(q^2) &= C_{10} f^+(q^2) , \notag \\
  \mathcal{L}_2(q^2) &= \frac{C_{10}}{2}
  \left[ f^+(q^2) \left( 1-\frac{m_B^2-m_K^2}{q^2} \right)
  +f^0(q^2)\frac{m_B^2-m_K^2}{q^2}
  \right].
\end{align}
The kaon's momentum $\bm p_K$ is in the rest frame of the $B$ meson and it is given by
\begin{equation}
  | \bm p_K| =
  \frac{\lambda^{1/2}(m_B^2,m_K^2,4m^2)}{2m_B},
\end{equation}
and $\lambda(x,y,z) = x^2+y^2+z^2-2xy-2xz-2yz$ is the K\"{a}llen function.
It is interesting to note that eq.~\eqref{eqn:rateBtoKspin1} and~\eqref{eqn:rateBtoKspin0} behave like $m^2$ in the $m\to 0$ limit.

The decay rates of $B\to K^*+$ leptonium are more complicated due to the presence of a larger number of form factors; they can be cast in the following form:
\begin{multline}
  \Gamma(B \to K^* + \text{ortho-leptonium}) =  
  \frac{G_F^2 |\pks|^3 m^2}{256 \pi^4}
  \alpha^5 \zeta(3) |V_{tb} V_{ts}^*|^2 
   \Bigg\{
  8 m^2 |\mathcal{M}_1|^2  \label{eqn:decayrateBtoKsspin1} \\ 
  + \frac{1}{\mks^2} \Bigg[
  |\mathcal{M}_2|^2
  \left(1+\frac{12 m^2 \mks^2}{m_B^2 |\pks|^2} \right) 
  + |\mathcal{M}_3|^2 m_B^2 |\pks|^2  
  -\mathrm{Re} \left( \mathcal{M}_2 \mathcal{M}_3^*\right)
  (m_B^2-\mks^2-4m^2)
\Bigg]
\Bigg\} \, , 
\end{multline}
\begin{multline}
  \Gamma(B \to K^* + \text{para-leptonium}) =
  \frac{G_F^2 |\pks|^3 m^2}{256 \pi^4 \mks^2}
  \alpha^5 \zeta(3) |V_{tb} V_{ts}^*|^2 
  \\
   \times
 \Bigg\{
   |\mathcal{N}_2|^2
   +\frac{|\mathcal{N}_3|^2}{4}
   \left( m_B^2-\mks^2-4m^2 \right)^2
   -\mathrm{Re}\left( \mathcal{N}_2\mathcal{N}_3^* \right)
 \left( m_B^2-\mks^2-4m^2 \right) \Bigg\} \, ,
 \label{eqn:decayrateBtoKsspin0}
\end{multline}
where $\mathcal{M}_i = \mathcal{M}_i(4m^2)$ and $\mathcal{N}_i = \mathcal{N}_i(4m^2)$.
The explicit expressions of the functions $\mathcal{M}_i(q^2)$ and $\mathcal{N}_i(q^2)$ are reported in the appendix.

The decay rate $\Gamma(B \to K^* + $ para-leptonium) behaves as  $m^2$ when $m\to0$, while $\Gamma(B \to K^* + $ ortho-leptonium) is approximately flat when $m\to0$: the $1/q^2$ poles appearing in $\mathcal{M}_{1,2,3}$ --- arising from the photon propagator with $C_{7\gamma}$ --- is canceled in eq.~\eqref{eqn:rateBtoKspin1} by an equal power of $m$ at the numerator.

The numerical values of the branching ratios are reported in table~\ref{tab:br}. For $\ell = e, \mu$ we took the values of the various form factors at $q^2=0$, while in the case of the tau they are evaluated at $q^2=4m_\tau^2$ employing the explicit numerical parametrization in~\cite{Khodjamirian:2010vf}. 
Moreover for $\ell=\tau$ the factor $\zeta(3)$ appearing in the decay rate is substituted with one, 
see the discussion at the end of section~\ref{sec:onium}.  Also, charm-loop effects are not taken into account; these will be discussed 
separately. 

The last column of table~\ref{tab:br} shows the peculiar dependence of  $\Gamma(B \to K^* + $ortho-leptonium) on the lepton 
mass: the branching ratios of $B \to K^* + $ ortho-positronium and $B\to K^* + $ ortho-muonium are almost equal. 
\begin{table}[h]
\def\arraystretch{1.2}
  \centering
  \begin{tabular}{crrrr}
    \toprule
    $\ell$ & $B\to K $ para &$ B\to K $ ortho
    & $B\to K^* $ para &$ B\to K^* $ ortho \\
    \hline
    $e$ &$2.3\times10^{-20}$ &$1.6\times 10^{-20}$ &
    $1.7\times 10^{-20}$ & $2.5\times 10^{-14}$ \\
    $\mu$ & $9.8\times 10^{-16}$ & $6.9 \times 10^{-16}$ &
    $7.2\times10^{-16}$& $2.5\times 10^{-14}$  \\
    $\tau$ &$2.9\times10^{-13}$ & $1.2\times10^{-13}$ &
    $1.3\times 10^{-13}$ & $2.5\times 10^{-13}$ \\ 
    \bottomrule
  \end{tabular}
  \caption{The branching ratios of $B\to K^{(*)}+$ leptonium stemming from the operators $O_{7\gamma},O_9$ and $O_{10}$.}
  \label{tab:br}
\end{table}

We note that some of the branching rations are too small even for high-luminosity experiments, however some of the branching ratios with muonia and tauonia may become accessible at the LHCb. 
The use of these decay modes is twofold. 
On the one hand it allows an independent probe for the Wilson coefficients, since the para-leptonia involve only the Wilson coefficient $C_{10}$ while the decays into ortho-leptonia will be sensitive to $C_9$. 
On the other hand, the decay into a tauonuim state will give us an independent  handle on the $b \to s \tau^+ \tau^-$ interaction, based on a different way to reconstruct it.   

\section{Charm-loop Effect}
\label{sec:charmloop}
So far we have estimated the dominant contribution to the branching ratios generated by the operators 
$O_{7\gamma}$, $O_9$ and $O_{10}$ in the effective Hamiltonian, leading to local hadronic current, which can be handled by 
form factors. In this section we will discuss the charm-loop effect in the $B\to K^{(*)} \ell^+\ell^-$ amplitude on the decay modes 
we consider here. 

This effect is generated by the current-current operators $O_1$ and $O_2$ acting together with the $c$-quark electromagnetic current, 
which eventually produces the leptonium through electromagnetic interaction. Perturbatively this is described by a 
charm-loop which couples to the lepton pair by means of a virtual photon.

However, the charm loop becomes a genuine long-distance hadronic effect if the lepton invariant mass $q^2$ is near the 
charmonium resonances or above the $D \overline{D}$ threshold. Consequently, this is not a problem for the light leptonia, 
since their masses are far below the $c\bar{c}$ threshold. However,  
the tau-pair threshold lies right between the $J/\psi$ and the $\psi(2S)$ resonances and the problem is much more severe for the tauonium. 

To leading order in $\alpha$ we only need to consider the case of an ortho-leptonium in the final state, 
which is produced by one off-shell photon coupled to the charm so that the amplitude has the same power 
of $\alpha$ as the amplitude with $O_{7\gamma}$ and $O_9$. 
The perturbative charm-loop contribution to the $B\to K^{(*)} \ell \ell$ amplitude can be conveniently expressed as a process and $q^2$ dependent correction to the Wilson coefficient $C_9$~\cite{Grinstein:1988me,Misiak:1992bc,Buras:1994dj}:
\begin{equation}
  C_9 \to C_9+\Delta C_9^{\bar c c}(q^2)
  \quad \mbox{with} \quad
  \Delta C_9^{\bar c c}(q^2) =
  (C_1+3C_2) h(m_c^2,q^2).
  \label{eqn:deltaC9}
\end{equation}
The function $h(m_c^2,q^2)$ comes from the one-loop matrix element of the four-quark operators $O_1$ and $O_2$ and has the form:
\begin{align}
  h(m_c^2,q^2) &=
  -\frac{8}{9}\log \left( \frac{m_c}{m_b} \right)
  +\frac{8}{27}
  +\frac{4}{9} y \notag \\
  &-\frac{2}{9} (2+y) |1-y|^{\frac{1}{2}}
  \begin{cases}
    \log \left( \frac{1+\sqrt{1-y}}{1-\sqrt{1-y}} \right) -i \pi  &
    \text{if } 0<y<1, \\
    2 \arctan \left( \frac{1}{\sqrt{y-1}} \right)  &
    \text{if } y>1,
  \end{cases}
\end{align}
where $y=4m_c^2/q^2$.  We note that this expression has an expansion in $q^2$, 
\begin{equation}
h(m_c^2,q^2) =  
-\frac{8}{9} \log \left( \frac{m_c}{m_b} \right) 
-\frac{4}{9} + \frac{4}{45}  \frac{q^2}{m_c^2} 
+ O \left( q^2/m_c^2  \right)^2 \, ,  
\end{equation} 
and thus, up to small effects of the order $4 m^2 / m_c^2$, the function $h$ yields for the light leptonia the shift
\begin{equation}
  C_9 \to C_9 - (C_1+3C_2) \left( \frac{8}{9} \log \left( \frac{m_c}{m_b} \right) +\frac{4}{9}  \right). 
\end{equation}
The shift in $C_9$ implies a correction $\delta_{\bar c c}$  defined as 
\begin{equation}
  \Gamma = \Gamma^{O_{7\gamma},O_9} (1+\delta_{\bar c c}) \, , 
\end{equation}
which is the same for  both $e$ and $\mu$ and 
which amounts to $ \sim 10\%$ for a $K$ in the final state, while it is a negligible contribution for a $B$ decaying into a $K^*$ since the $B\to K^*+$ leptonium decay is dominated almost entirely by $O_{7\gamma}$ when $q^2 \ll m_B^2$. 
The result is shown in table~\ref{tab:brcc}. 
 
\begin{table}[h]
\def\arraystretch{1.2}
  \centering
  \begin{tabular}{ccccc}
    \toprule
     &\multicolumn{2}{c}{$ B\to K $ ortho} & \multicolumn{2}{c}{$ B\to K^* $ ortho} \\
    $\ell$ & $\delta_{\bar c c}$ & Br &   $\delta_{\bar c c}$ & Br\\
    \hline
    $e$ & 0.11 & $1.8\times 10^{-20}$ & $\sim0$ & $2.5\times 10^{-14}$ \\
    $\mu$ & 0.11 & $7.7 \times 10^{-16}$ & $\sim0$ & $2.5\times 10^{-14}$ \\
    \bottomrule
  \end{tabular}
  \caption{The branching ratios of $B\to K^{(*)}  +$ positronium and $B\to K^{(*)}   + $ muonium  inclusive of the leading 
   perturbative charm-loop effect arising from the operators $O_1$ and  $O_2$.  }
  \label{tab:brcc}
\end{table}

For the tauonium we first look at the perturbative result close to threshold, where we find 
\begin{equation} \label{ThresExp}
h(m_c^2,q^2) =  -\frac{8}{9} \log \left( \frac{m_c}{m_b} \right) 
+\frac{20}{27}
- \frac{16}{9} \left( \frac{q^2-4m_c^2}{ 4 m_c^2}    \right)
 + O \left( (q^2 - 4 m_c^2) / (4m_c^2)  \right)^2 \, .  
\end{equation} 
Since we have $q^2 = 4 m_\tau^2 \sim 4 m_c^2$ the major part is the constant term which is numerically close to  $h(m_c^2,0)$ 
and hence we expect an effect close to the one for the light leptonia. In table~\ref{tab:brcc2} we show the corresponding results, where in the calculation we inserted the full expression of $h(m_c^2,q^2)$. 
 
\begin{table}[h]
\def\arraystretch{1.2}
  \centering
  \begin{tabular}{ccccccc}
    \toprule
     &\multicolumn{3}{c}{$ B\to K $ ortho} & 
     \multicolumn{3}{c}{$ B\to K^* $ ortho} \\
     $\ell$ &
     $\delta_{\bar c c} $ & Br &   
     $\delta_{\bar c c}$ &  Br \\
    \hline
    $\tau$ &
    0.18 &  $1.4\times10^{-13}$ &
    0.21 &  $3.0\times 10^{-13}$ \\
    \bottomrule
  \end{tabular}
  \caption{The branching ratios of $B\to K^{(*)} +$ tauonium inclusive of the leading perturbative charm-loop effects arising from the operators $O_1$ and  $O_2$.  }
  \label{tab:brcc2}
\end{table}

However, close to the threshold the real spectral function can be quite different from the partonic result, and thus the 
approximation to keep only the constant term in (\ref{ThresExp}) is much less reliable as in the case for the light leptonia. 
In order to get some idea if we should expect large corrections, 
we try to naively estimate the non-perturbative effects in tauonium production by considering a single resonance exchange. 
Since the $\psi(2S)$  can decay into two $\tau$ leptons with a known branching fraction, we consider  
the $B\to K^{(*)} \psi (2S)$ decay, followed by the subsequent mixing of the $\psi$ into a $\tau^+\tau^-$ bound state. 
We can express the width of this subsequent  process as:
\begin{equation}
  \Gamma(B\to K^{(*)}(\psi \to \text{tauonium})) =
  \Gamma(B\to K^{(*)}\psi) \, \text{Br}(\psi \to \tau^+\tau^-)
  \, \delta_\tau,
  \label{eqn:BtoKpsitota}
\end{equation}
where $\delta_\tau$ is a correction that allows us to compare the experimental value of the  $\psi(2S)\to \tau^+ \tau^-$ branching ratio with the case where a tauonium shows up in the final state 
\begin{equation}
  \delta_\tau = \frac{\Gamma(\gamma^* \to \text{tauonium})}{\Gamma(\gamma^* \to \tau^+\tau^-)}.
\end{equation}
The factor $\delta_\tau$ contains the squared wave function of the tauonium and some phase space factor. 
By comparing the expression of the tauonium matrix element in eq.~\eqref{eqn:matelemvector}  and 
the $\gamma^*\to \tau^+\tau^-$ amplitude in the non relativistic approximation in eq.~\eqref{eqn:matV}, we obtain:
\begin{equation}
  \delta_\tau = 8 \pi^2 |\phi_{n,\bm 0}(0)|^2 \frac{\sqrt{q^2}}{m p_\tau} \delta(q^2-M_n^2),
\end{equation}
where $M_n \sim 2m_\tau$ is the tauonium mass and $p_\tau = \sqrt{q^2-4m_\tau^2}/2$.
The photon momentum $q$ appearing in $\delta_\tau$ must be evaluated at $q^2=m_\psi^2$. The $\delta$ function 
appearing in the expression for $\delta_\tau$ reflects the fact that we have treated the $\psi(2S)$ as a stable particle; 
including a finite width for this state means that we replace
the $\delta$-function by a Bright-Wigner distribution:
\begin{equation}
  \pi \delta(m_\psi^2-4m_\tau^2) \to
  \frac{\Gamma_\psi m_\psi}{(m_\psi^2-4m^2_\tau)^2 +\Gamma_\psi^2m_\psi^2}.
\end{equation}
With such approximation, we can insert the explicit expression of $\delta_\tau$ back in eq.~\eqref{eqn:BtoKpsitota} to obtain
\begin{eqnarray}
&&  \Gamma(B\to K^{(*)}(\psi \to \text{tauonium})) =  \nonumber \\ 
&& \qquad   \Gamma(B\to K^{(*)}\psi) \, \Gamma(\psi \to e^+e^-)
  \frac{2 \alpha^3 m_\tau^2 m_\psi}
   {(m_\psi^2-4m^2_\tau)^2 +\Gamma_\psi^2m_\psi^2}
   \left( 1+\frac{2m_\tau^2}{m_\psi^2} \right) \, , 
   \label{eqn:BtoKTares}
\end{eqnarray}
where we have written the $\psi \to \tau \tau$ branching ratio in terms of the $\psi \to e^+ e^-$ one:
\begin{equation}
  \frac{ \Gamma(\psi \to \tau^+ \tau^-)}{\Gamma(\psi \to e^+ e^-)} =
  \frac{2 p_\tau}{m_\psi} 
  \left( 1+\frac{2m_\tau^2}{m_\psi^2} \right).
\end{equation}
Using as an input  in~\eqref{eqn:BtoKTares} the values of the branching ratios 
Br$(B\to K^* \psi )=(6.32 \pm 0.37) \times 10^{-4}$,  Br$(B\to K^* \psi )=(5.92 \pm 1.23) \times 10^{-4}$~\cite{Amhis:2016xyh} and Br$(\psi(2S)\to e^+e^-) = (7.89 \pm 0.17)\times10^{-3}$~\cite{Patrignani:2016xqp}, we obtain 
\begin{align}
  \text{Br}(B\to K (\psi \to \text{tauonium})) &=
   2.1 \times 10^{-14}, \\
   \text{Br}(B\to K^* (\psi \to \text{tauonium})) &=
  2.0 \times 10^{-14} \, , 
\end{align}
yielding indeed a $18\%$ and a $8\%$ correction to the branching ratios given by $O_{7\gamma}$ and $O_9$ in table~\ref{tab:br},respectively. 
We take this as an indication that the corrections obtained form the perturbative reasoning will not exceed 20\% also for the case of the tauonium.

\section{Conclusions}
\label{sec:conclusions}
We have presented a calculation of the branching ratios for the decays $B \to K^{(*)}+$ leptonium. The predictions for these branching fractions turn out to be quite precise; for the light leptonia the branching fractions can be predicted at the level of 5\%, while for the tauonium-final state the uncertainties are a bit larger. 

As expected from the naive counting of parameters, these branching fractions are very small, but some of the decays with muon and tau in the final states will be within the reach of the LHCb with an integrated luminosity of $50$ fb$^{-1}$. 

The $\ell^+\ell^-$ bound state systems have a clear signature for their  decays and  it is possible to distinguish between the two spin states ortho- and para-leptonium. 
 The advantage is twofold: these decays can give first an independent analysis of the helicity structure of the underlying $b \to s \ell \ell$ interaction, where the decays into ortho-leptonia test different structures as the decays into  para-leptonia.
Secondly, the decays $B \to K^{(*)}+$ tauonium do not require a reconstruction of the tau leptons. This can allow us a cross check of the $B \to K^{(*)} \tau^+ \tau^-$ results, which require a reconstruction of the two final state leptons, by comparing them to the $B \to K^{(*)}+$ tauonium decays.

\section*{Acknowledgements}
We thank Alexander Khodjamirian for useful discussions and comments. 
This work was supported by DFG through the Research Unit FOR 1873 ``Quark Flavour Physics and Effective Field Theories''.

\appendix

\section{Effective Hamiltonian}
The effective Hamiltonian for the $\Delta B= \Delta S = 1$ decays is defined to be
\begin{equation}
  H_{\mysmall eff} =
  -\frac{4 G_F}{\sqrt{2}} V_{tb} V_{ts}^* \sum_{i=1}^{10} C_i(\mu) O_i(\mu).
\end{equation}
The operators $O_i$ whose Wilson coefficients enter in eqs.~(\ref{eqn:rateBtoKspin1}-\ref{eqn:decayrateBtoKsspin0}) and~\eqref{eqn:deltaC9} are
\begin{align}
  O_1 &=(\bar{s} \gamma^\mu P_L c) (\bar{c} \gamma_\mu P_L b) &
  O_2 &=(\bar{s}^\alpha \gamma^\mu P_L c^\beta) (\bar{c}^\beta \gamma_\mu P_L b^\alpha) \notag \\
  O_9 &=\frac{ \alpha}{4\pi} 
  (\bar{s} \gamma^\mu P_L b)(\bar \ell \gamma_\mu \ell), &
  O_{7\gamma} &=
  \frac{e}{16 \pi^2} 
  \bar s \sigma^{\mu\nu}( m_s P_L+m_b P_R) b \, F_{\mu\nu}, \notag\\
  O_{10} &=\frac{ \alpha}{4\pi} 
  (\bar{s} \gamma^\mu P_L b)(\bar \ell \gamma_\mu \gamma_5 \ell) ,
\end{align}
where $P_L = (1-\gamma_5)/2$ and $P_R=(1+\gamma_5)/2$. We adopted the approximation $|V_{tb}V^*_{ts}| \simeq |V_{cb} V^*_{cs}| = 0.040$~\cite{Charles:2004jd,Patrignani:2016xqp}.  
We used the Wilson coefficients calculated with LO running; their numerical values are: $C_1(\bar m_b)=1.12, C_2(\bar m_b)=-0.27, C_7^{\mysmall eff}(\bar m_b) = C_{7\gamma}-\frac{1}{3}C_5-C_6 = -0.32, C_9(\bar m_b)=4.2, C_{10}(\bar m_b) = -4.4$.

\section{Decay amplitudes and form factors}
The amplitude of $B\to K \ell^+ \ell^-$ given by the hadronic matrix element of the operators $O_{7\gamma}, O_9 $ and $O_{10}$ is
\begin{align}
  \mathcal{A} (B \to K \ell^{+} \ell^{-}) &=
  {G_F \over  \sqrt{2} }
  \frac{\alpha }{ \pi} V_{tb} V_{ts}^* \Bigg\{
    \bar{\ell}\gamma_{\mu} \ell\, p^\mu\bigg( C_9 f^{+}(q^2)
     + {2 (m_b+m_s) \over m_B+m_K} C_7^{\mysmall eff} f^{T}(q^2) \bigg)
     \notag \\
     &+ \bar{\ell} \gamma_{\mu} \gamma_5 \ell \,
       p^\mu C_{10}f^{+}(q^2)
     \notag \\
       &+ \bar{\ell} \gamma_{\mu} \gamma_5 \ell \,q^\mu 
       \frac{C_{10}}{2}
       \Bigg[
	 f^+(q^2) \left( 1-\frac{m_B^2-m_K^2}{q^2} \right)
	 +f^0(q^2)\frac{m_B^2-m_K^2}{q^2}
       \Bigg]
     \Bigg\} \,. 
\end{align}
The $B\to K$ form factors are defined as
\begin{align}
\langle K(p)|\bar{s} \gamma_\mu b|B(p+q)\rangle &=
f^+(q^2) \left[ 2 p_\mu + 
  \left( 1-\frac{m_B^2-m_K^2}{q^2} \right) q_\mu
\right] 
+f^0(q^2)\frac{m_B^2-m_K^2}{q^2}q_\mu, \\
\langle K(p)|\bar{s} \sigma_{\mu \rho} q^{\rho}b|B(p+q)\rangle &=
\bigg[q^2(2p_\mu +q_\mu) -(m_B^2-m_K^2)q_\mu\bigg]
\frac{i f^{T}(q^2)}{m_B+m_K}\,.
\end{align}
The amplitude for the $B\to K^* \ell^+ \ell^-$ is given by the following expression~\cite{Khodjamirian:2010vf}:

\begin{align}
  \mathcal{A} (B \to K^{\ast} l^{+} l^{-}) &= \notag \\
  { G_F \over 2 \sqrt{2} }
 {\alpha \over \pi} V_{tb} V_{ts}^{\ast} 
 &
 \Bigg\{ \bar{l} \gamma^{\mu} l 
 \Big[
   \epsilon_{ \mu \nu \rho \sigma }
  \epsilon^{\ast \nu} q^{\rho} p^{\sigma} {\cal M}_1(q^2)
  -i \epsilon^{\ast}_{\mu} {\cal M}_2(q^2) 
  + i  (\epsilon^{\ast} \cdot q)p_{\mu}{\cal M}_3(q^2)
\Big] \notag \\
& +\bar{l} \gamma^{\mu} \gamma_5 l\Big[\epsilon_{ \mu \nu \rho
\sigma } \epsilon^{\ast \nu} q^{\rho} p^{\sigma} {\cal N}_1(q^2)
-i \epsilon^{\ast}_{\mu} {\cal N}_2(q^2)
+ i  (\epsilon^{\ast} \cdot q)p_{\mu}{\cal N}_3(q^2) \Big] \Bigg\}
\,, \hspace{1 cm}
\end{align}
where
\begin{eqnarray}
 {\cal M}_1(q^2) &=& C_9 { 2 V(q^2)\over
  m_B+m_{K^{\ast}}} + 4 C_7^{\mysmall eff} { m_b + m_s \over q^2} T_1
  (q^2)\,,
  \nonumber \\
 {\cal M}_2(q^2)&=& C_9 (m_B+m_{K^{\ast}}) A_1(q^2)
  \nonumber \\ && +  2 C_7^{\mysmall eff} (m_B^2 - m_{K^{\ast}}^2) { m_b +
    m_s \over q^2} T_2(q^2)\,,
  \nonumber\\
  {\cal M}_3(q^2) &=&
  2 C_9 { A_2(q^2)  \over m_B+m_{K^{\ast}} }
   \nonumber\\
   && + 4 C_7^{\mysmall eff} { m_b - m_s \over q^2} \bigg(T_2(q^2) 
   + {q^2 \over m_B^2 - m_{K^{\ast}}^2}
T_3(q^2) \bigg) \,, 
\end{eqnarray}
 and
 \begin{eqnarray}
	{\cal N}_1(q^2) &=& 2 C_{10} {V(q^2) \over m_B+
	m_{K^{\ast}}},
	\nonumber \\
	{\cal N}_2(q^2) &=& C_{10} ({m_B +
	  m_{K^{\ast}}}) A_1(q^2)\,,
	  \nonumber \\
	  {\cal N}_3(q^2)&=&
	   2 C_{10}  { A_2(q^2) \over m_B +
	   m_{K^{\ast}}}\, .
\end{eqnarray}
In this paper we employed for the $B\to K^{(*)}$ form factors the $q^2$-parametrization introduced in ref.~\cite{Bourrely:2008za}.  The form factors values at $q^2=0$ and the slope parameters calculated from light-cone sum rules can be found in the appendix of ref.~\cite{Khodjamirian:2010vf}.
The $B \to K^{\ast}$ form factors are defined as
\begin{equation}
  \langle K^{\ast}(p)|\bar{s} \gamma^\mu b|B(p+q)\rangle =
  V(q^2) \epsilon^{\mu\sigma}\, _{\nu\rho} \,
  \epsilon_\sigma^*
  \frac{2(p+q)^\nu p^\rho}{m_B+\mks}, 
\end{equation}
\begin{align}
  \langle K^{\ast}(p)|\bar{s} \gamma^\mu \gamma_5 b|B(p+q)\rangle &=
i \epsilon^*_\nu \Bigg[
  2 \mks A_0(q^2) \frac{q^\mu q^\nu}{q^2}
  +A_1(q^2) (m_B+\mks) \eta^{\mu\nu}  \notag \\
  &-A_2(q^2) \eta^{\mu\sigma} 
  \frac{(q+2p)_\sigma q^\nu}{m_B+\mks}
\Bigg],
  \\
  \langle K^{\ast}(p)|\bar{s} \sigma_{\mu \rho} q^{\rho} (1+
	   \gamma_5) b|B(p+q)\rangle &= 
	   2 i \epsilon_{\mu \nu \rho \sigma }
	   \epsilon^{\ast \nu } q^{\rho}  p^{\sigma}
	   T_1(q^2)
	   \notag \\
	   &+ [(m_B^2- m^2_{K^{\ast}})
	     {\epsilon}^{\ast}_\mu -(\epsilon^\ast \!\cdot q) 
	   (2p+q)_{\mu}] \, T_2(q^2)
	   \notag \\
	   &+  (\epsilon^\ast\! \cdot q)
	   \bigg[q_{\mu} - {q^2 \over m_B^2- m^2_{K^{\ast}}}
	   (2 p+q)_{\mu}\bigg] T_3(q^2) \, ,
\end{align}
where $\epsilon_\mu$ is the polarization vector of the vector meson and $\eta^{\mu\nu} =g^{\mu\nu}-q^\mu q^\nu/q^2$.


\footnotesize

\begin{thebibliography}{99}
\bibitem{Aaij:2014ora}
  R.~Aaij {\it et al.} [LHCb Collaboration],
  Phys.\ Rev.\ Lett.\  {\bf 113} (2014) 151601
  [arXiv:1406.6482 [hep-ex]].

\bibitem{Aaij:2017vbb}
  R.~Aaij {\it et al.} [LHCb Collaboration],
  JHEP {\bf 1708} (2017) 055
  [arXiv:1705.05802 [hep-ex]].

\bibitem{Capdevila:2017bsm} 
  B.~Capdevila, A.~Crivellin, S.~Descotes-Genon, J.~Matias and J.~Virto,
  JHEP {\bf 1801}, 093 (2018)
  [arXiv:1704.05340 [hep-ph]].

\bibitem{Altmannshofer:2017yso}
  W.~Altmannshofer, P.~Stangl and D.~M.~Straub,
  Phys.\ Rev.\ D {\bf 96} (2017) no.5,  055008
  [arXiv:1704.05435 [hep-ph]].

\bibitem{Geng:2017svp}
  L.~S.~Geng, B.~Grinstein, S.~Jäger, J.~Martin Camalich, X.~L.~Ren and R.~X.~Shi,
  Phys.\ Rev.\ D {\bf 96} (2017) no.9,  093006
  [arXiv:1704.05446 [hep-ph]].

\bibitem{Ciuchini:2017mik}
  M.~Ciuchini, A.~M.~Coutinho, M.~Fedele, E.~Franco, A.~Paul, L.~Silvestrini and M.~Valli,
  Eur.\ Phys.\ J.\ C {\bf 77} (2017) no.10,  688
  [arXiv:1704.05447 [hep-ph]].

\bibitem{Bordone:2016gaq}
  M.~Bordone, G.~Isidori and A.~Pattori,
  Eur.\ Phys.\ J.\ C {\bf 76} (2016) no.8,  440
  [arXiv:1605.07633 [hep-ph]].

\bibitem{Alves:2008zz}
  A.~A.~Alves, Jr. {\it et al.} [LHCb Collaboration],
  JINST {\bf 3} (2008) S08005.

\bibitem{Deutsch:1951zza}
  M.~Deutsch,
  Phys.\ Rev.\  {\bf 82} (1951) 455.

\bibitem{Bogomyagkov:2017uul}
  A.~Bogomyagkov, V.~Druzhinin, E.~Levichev, A.~Milstein and S.~Sinyatkin,
  arXiv:1708.05819 [physics.acc-ph].

\bibitem{Itzykson:1980rh}
  C.~Itzykson and J.~B.~Zuber,
  ``Quantum Field Theory,''
  Mcgraw-hill (1980).

\bibitem{Berestetsky:1982aq}
  V.~B.~Berestetskii, E.~M.~Lifshitz and L.~P.~Pitaevskii,
  ``Quantum Electrodynamics,''
   Butterworth-Heinemann (1982).

\bibitem{Kniehl:2009pg}
  B.~A.~Kniehl, A.~V.~Kotikov and O.~L.~Veretin,
  Phys.\ Rev.\ A {\bf 80} (2009) 052501
  [arXiv:0909.1431 [hep-ph]].

\bibitem{Adkins:2015wya}
  G.~S.~Adkins,
  Hyperfine Interact.\  {\bf 233} (2015) no.1-3,  59.

\bibitem{Adkins:2003eh}
  G.~S.~Adkins, N.~M.~McGovern, R.~N.~Fell and J.~Sapirstein,
  Phys.\ Rev.\ A {\bf 68} (2003) 032512
  [hep-ph/0305251].

\bibitem{Adkins:2001zz}
  G.~S.~Adkins, R.~N.~Fell and J.~Sapirstein,
  Phys.\ Rev.\ A {\bf 63} (2001) 032511.

\bibitem{Adkins:2000fg}
  G.~S.~Adkins, R.~N.~Fell and J.~R.~Sapirstein,
  Phys.\ Rev.\ Lett.\  {\bf 84} (2000) 5086
  [hep-ph/0003028].

\bibitem{Penin:2003jz}
  A.~A.~Penin,
  Int.\ J.\ Mod.\ Phys.\ A {\bf 19} (2004) 3897
  [hep-ph/0308204].

\bibitem{Kniehl:2000dh}
  B.~A.~Kniehl and A.~A.~Penin,
  Phys.\ Rev.\ Lett.\  {\bf 85} (2000) 1210
   Erratum: [Phys.\ Rev.\ Lett.\  {\bf 85} (2000) 3065]
  [hep-ph/0004267].

\bibitem{Karshenboim:2005iy}
  S.~G.~Karshenboim,
  Phys.\ Rept.\  {\bf 422} (2005) 1
  [hep-ph/0509010].
  
\bibitem{Hill:2000qi}
  R.~J.~Hill and G.~P.~Lepage,
  Phys.\ Rev.\ D {\bf 62} (2000) 111301
  [hep-ph/0003277].

\bibitem{Melnikov:2000fi}
  K.~Melnikov and A.~Yelkhovsky,
  Phys.\ Rev.\ D {\bf 62} (2000) 116003
  [hep-ph/0008099].

\bibitem{Czarnecki:1999gv}
  A.~Czarnecki, K.~Melnikov and A.~Yelkhovsky,
  Phys.\ Rev.\ Lett.\  {\bf 83} (1999) 1135
  [hep-ph/9904478].

\bibitem{Czarnecki:1999ci}
  A.~Czarnecki, K.~Melnikov and A.~Yelkhovsky,
  Phys.\ Rev.\ A {\bf 61} (2000) 052502
  [hep-ph/9910488].

\bibitem{Malenfant:1987tm}
  J.~Malenfant,
  Phys.\ Rev.\ D {\bf 36} (1987) 863.

\bibitem{Jentschura:1997tv}
  U.~D.~Jentschura, G.~Soff, V.~G.~Ivanov and S.~G.~Karshenboim,
  Phys.\ Rev.\ A {\bf 56} (1997) 4483
  [physics/9706026].

\bibitem{Karshenboim:1998we}
  U.~D.~Jentschura, S.~G.~Karshenboim, V.~G.~Ivanov and G.~Soff,
  Phys.\ Lett.\ B {\bf 424} (1998) 397
  [hep-ph/9706401].

\bibitem{Perl:1993sk}
  M.~L.~Perl,
  In ``Stanford 1992, The third family and the physics of flavor'' 213-252


\bibitem{Novikov:1977dq}
  V.~A.~Novikov, L.~B.~Okun, M.~A.~Shifman, A.~I.~Vainshtein, M.~B.~Voloshin and V.~I.~Zakharov,
  Phys.\ Rept.\  {\bf 41} (1978) 1.   

\bibitem{Pivovarov:1991vb}
  A.~A.~Pivovarov,
  Phys.\ Rev.\ D {\bf 47} (1993) 5183.

\bibitem{Khodjamirian:2010vf} 
  A.~Khodjamirian, T.~Mannel, A.~A.~Pivovarov and Y.-M.~Wang,
  JHEP {\bf 1009}, 089 (2010)
  [arXiv:1006.4945 [hep-ph]].

\bibitem{Grinstein:1988me}
  B.~Grinstein, M.~J.~Savage and M.~B.~Wise,
  Nucl.\ Phys.\ B {\bf 319} (1989) 271.

\bibitem{Misiak:1992bc}
  M.~Misiak,
  Nucl.\ Phys.\ B {\bf 393} (1993) 23
   Erratum: [Nucl.\ Phys.\ B {\bf 439} (1995) 461].

\bibitem{Buras:1994dj}
  A.~J.~Buras and M.~Munz,
  Phys.\ Rev.\ D {\bf 52} (1995) 186
  [hep-ph/9501281].

\bibitem{Amhis:2016xyh}
  Y.~Amhis {\it et al.} [HFLAV Collaboration],
  Eur.\ Phys.\ J.\ C {\bf 77} (2017) no.12,  895
  [arXiv:1612.07233 [hep-ex]].

\bibitem{Patrignani:2016xqp}
  C.~Patrignani {\it et al.} [Particle Data Group],
  Chin.\ Phys.\ C {\bf 40} (2016) no.10,  100001.

\bibitem{Buchalla:1995vs}
       G.~Buchalla, A.~J.~Buras and M.~E.~Lautenbacher,
         Rev.\ Mod.\ Phys.\  {\bf 68} (1996) 1125
	 [hep-ph/9512380].

\bibitem{Charles:2004jd}
  J.~Charles {\it et al.} [CKMfitter Group],
  Eur.\ Phys.\ J.\ C {\bf 41} (2005) no.1,  1
  [hep-ph/0406184].

\bibitem{Bourrely:2008za}
  C.~Bourrely, I.~Caprini and L.~Lellouch,
  Phys.\ Rev.\ D {\bf 79} (2009) 013008
   Erratum: [Phys.\ Rev.\ D {\bf 82} (2010) 099902]
  [arXiv:0807.2722 [hep-ph]].
\end{thebibliography}
\end{document}